# Dual-isotope narrow-line MOT of dysprosium by phase modulation

M. Dürbeck[1], L. Reihs[1], J. Seifert[1], B. Choudhari[1], J.P. Marulanda-Serna[1], N. Werum[1], M. De Pas[1], G. Meijer[1], and G. Valtolina[1,2 *]

[1] Fritz-Haber-Institut der Max-Planck-Gesellschaft, Faradayweg 4-6, 14195 Berlin, Germany.
[2] Humboldt Universität zu Berlin, Newtonstrasse 15, 12489, Berlin

*Author to whom any correspondence should be addressed.

E-mail: valtolina@fhi-berlin.mpg.de

## Abstract

We report on the characterization of a narrow-line magneto-optical trap (MOT), where two different isotopes of the dysprosium (Dy) atom can be simultaneously loaded and trapped. We rely on phase modulation via high-power, free-space electro-optic modulators (EOMs) to generate optical sidebands at the correct detuning for two different isotopes. This technique is applied on the slowing transition at 421 nm and the MOT transition at 626 nm. Using modular resonant electronic circuits, we match the sideband separation to the isotope shift among the two target isotopes, realizing mixtures of the bosonic isotopes $^{164}$Dy -$^{162}$Dy and $^{162}$Dy -$^{160}$Dy. We exploit the interplay between radiation forces and gravity in the narrow-line MOT to control the spatial position of the two isotopes with the EOM drive frequency. We demonstrate full control of the isotopic mixture with the EOM drive, offering new and cost-effective capabilities for controlling mixtures of atomic species that can be trapped in a narrow-line MOT.

## 1. Introduction

Laser-cooled gases are an ideal platform for testing fundamental physics and realizing quantum degenerate gases[1]. Precise control over multiple optical frequencies is crucial for establishing robust optical cycling conditions for efficient laser cooling. In this context, spectral engineering using electro-optic modulators (EOMs) has attracted significant interest, particularly for molecular laser cooling. Molecules require several frequency components just to address their hyperfine structure and make laser cooling possible. These frequencies are typically spaced by tens of MHz to few GHz. Recently phase modulation techniques emerged as a robust method to synthesize different frequency components[2,3].
In this work, we exploit optical engineering by phase modulation to develop a cooling scheme that simultaneously traps two different isotopes of the Dy atom in a dual-isotope magneto-optical trap (DI-MOT).
Dy is a lanthanide atom with a large magnetic dipole moment of nearly 10 $\mu_B$. Dy's magnetic dipole moment stems from its open, submerged *4f*-shell of electrons, which results in a large orbital angular momentum $J$ = 8 in its electronic ground state. Among its stable isotopes, the most abundant and commonly studied are the bosonic isotopes $^{160}$Dy, $^{162}$Dy, and $^{164}$Dy, and the fermionic isotopes $^{161}$Dy and $^{163}$Dy[4]. Despite its rather complex electronic structure, Dy can be efficiently laser cooled, trapped, and further cooled to the quantum degenerate regime[5]. Together with erbium (Er), another strongly magnetic lanthanide species, Dy has become a central platform for the investigation of new many-body regimes, such as dipolar supersolidity[4,6–9]. The exploration of dipolar supersolids has been so far restricted to single-isotope quantum gases, primarily using $^{162}$Dy, $^{164}$Dy, and $^{166}$Er. A DI-MOT of Dy is the starting point for the realization of new dipolar mixtures[10], where more exotic supersolid phases have been predicted[11–13].

DI-MOTs have so far been realized for several atomic species[14–19]. The key challenge in creating a DI-MOT lies in providing multiple laser frequencies to address multiple isotopes. Such experiments typically rely either on using two independent laser sources, one for each isotope, or on sequentially cooling and trapping one isotope, storing it away, and then shifting the laser frequency to address the second isotope[20]. Here instead, we exploit the relatively simple cooling scheme of Dy, where no repumping light is required. We use EOMs to effectively duplicate the cooling light and simultaneously generate optical sidebands at the optimal detuning $\delta$ for two Dy isotopes. Previously a rubidium DI-MOT has been realized placing fibre-integrated EOMs between the low-power seed and the optical amplifying stage[21]. Instead, we demonstrate a more widely applicable scheme, where optical sidebands are directly imprinted onto the output of high-power lasers. This approach is necessary for the main optical transitions of Dy, where a seed-plus-amplifier architecture is not available. Our method therefore provides a cost-effective and dead-time-reducing solution for laser-cooling isotopic mixtures. It can be readily extended to different isotopic combinations and to other atomic species that benefit from repumper-free magneto-optical trapping, such as lanthanide and alkaline-earth atoms. In addition, since the cooling force is entirely realized by EOM-sidebands, we show a new isotope-dependent control of atoms in narrow-line MOTs, via fine tuning of the sidebands relative spacing, offering new tools for the control of mixed-species quantum computing and metrology platforms.

## 2. Implementation

Our experiment starts with a beam of hot Dy atoms produced in an effusion cell operating at 1050 °C. The Dy atoms are first slowed by a combination of transverse cooling (TC), Zeeman slowing, and Angled Slowing (AS) on the broad optical transition at 421 nm ($\Gamma_{421}= 2\pi \cdot 32.2$ MHz, Fig. 1a). We produce up to 2W of laser light at 421 nm using a frequency-doubled Ti:Sapphire laser. In the science chamber (Fig. 1b), the atoms are captured in a narrow-line MOT operating on the 626-nm transition ($\Gamma_{421}= 2\pi \cdot 135$ kHz, Fig.1a). To address the narrow-line 626-nm transition, we use a 1.5-W fiber laser system, which is frequency stabilized to the *a1* hyper-fine component of the *R86 (8-3)* ro-vibrational line of $^{127}I_2$ through polarisation spectroscopy[22]. For each laser system, we use an optical scheme based on acousto-optic modulators (AOMs) in order to shift the frequencies of the different beams. We initially optimize the 421- and 626-nm laser systems to laser cool a single Dy isotope, similarly to other experiments[23,24]. The MOT parameters are extracted by absorption imaging on the 421-nm transition. In the single-isotope configuration, we typically trap around $5 \cdot 10^7$ atoms in 5 seconds. A compression sequence of 200 ms cools the atoms to a temperature of 10 µK. At the end of this compression, we set the laser detuning $\delta/2\pi = -1.5$ MHz and we add an offset magnetic field $B_z = 2.1$ G along the gravity axis. $B_z$ is generated by a pair of coils in Helmholtz configuration and sets the equilibrium position of the compressed MOT.
Fig. 1c summarises the general idea behind our dual-isotope strategy. We place an EOM along each laser path and drive it at a frequency $\Omega = \frac{\Delta_{IS}}{2}$, where $\Delta_{IS}$ is the isotope shift among the selected isotopes at a specific optical transition. After the laser lock point has been moved by the drive frequency $\Omega$, we use the two first-order sidebands, separated by $\Delta_{IS}$, to simultaneously address the two target isotopes. The EOMs allow us to effectively duplicate our cooling and slowing light, creating laser beams at the appropriate detuning for two isotopes, which can then be loaded into the DI-MOT.
We initially test our scheme on the mixture composed of $^{164}$Dy and $^{162}$Dy, the two most abundant bosonic isotopes. The isotope shift $\Delta_{IS}$ between $^{164}$Dy and $^{162}$Dy for both the 421-nm and the 626-nm transition is slightly less than 1 GHz[25,26], falling within the bandwidth of commercial free-space EOMs. The drawback in using free-space EOMs is the large driving voltage required to achieve sufficient phase modulation depth $\beta$, to transfer sufficient laser power to the first-order sidebands (see Fig. 1f). Typically, free-space EOMs have built-in resonant *LC* circuits, which make the radio-frequency (RF) power requirements experimentally practical. However, *LC* circuits significantly reduce the EOM bandwidth to less than few tens of MHz. When switching to a different isotopic mixture, the limited tuning range of the internal resonant circuit may prevent our scheme to function at all. To circumvent this limitation, we use high-bandwidth non-resonant EOM and combine it with simple plug-and-play electronics, added at the EOM input, to induce a resonance in the circuit (see Fig. 1d). In this way, we significantly increase the modulation depth at a given RF power. Furthermore, changing the resonance frequency of the whole circuit allows us to target a different isotopic mixture.
We first test the modular electronics on the 626-nm laser scheme, where we use a non-resonant, traveling-wave EOM (model QUBIG TWP2M2-VIS, Fig. 1e). Since non-resonant EOMs can be modelled as simple capacitors *C*, we add an external inductor *L* at the EOM input port to improve the EOM quality factor *Q*. The isotope shift at 626 nm between $^{164}$Dy and $^{162}$Dy is $\Delta_{IS}$ = 963(3) MHz[26]. We use an inductance of 160 nH to realize a resonance at 481.6 MHz. Assuming a lumped capacitor model, we measure with a network analyser a *Q*-factor of approximately 130. However, when using a Fabry-Perot cavity to measure the effective sidebands (see Fig. 1f), we only measure a factor-of-three increase in sideband power at the same drive power. A likely explanation is that the traveling-wave configuration of the EOM deviates from the lumped-capacitor model and instead behaves as a distributed transmission line. In such a geometry, the assumption that the Q-factor directly corresponds to the voltage magnification no longer holds. Despite this, an RF power lower than 2.5 W is sufficient for maxing out the first-orders. At this RF-power, we do not observe any drift in the sideband intensity even at optical powers exceeding 500mW.

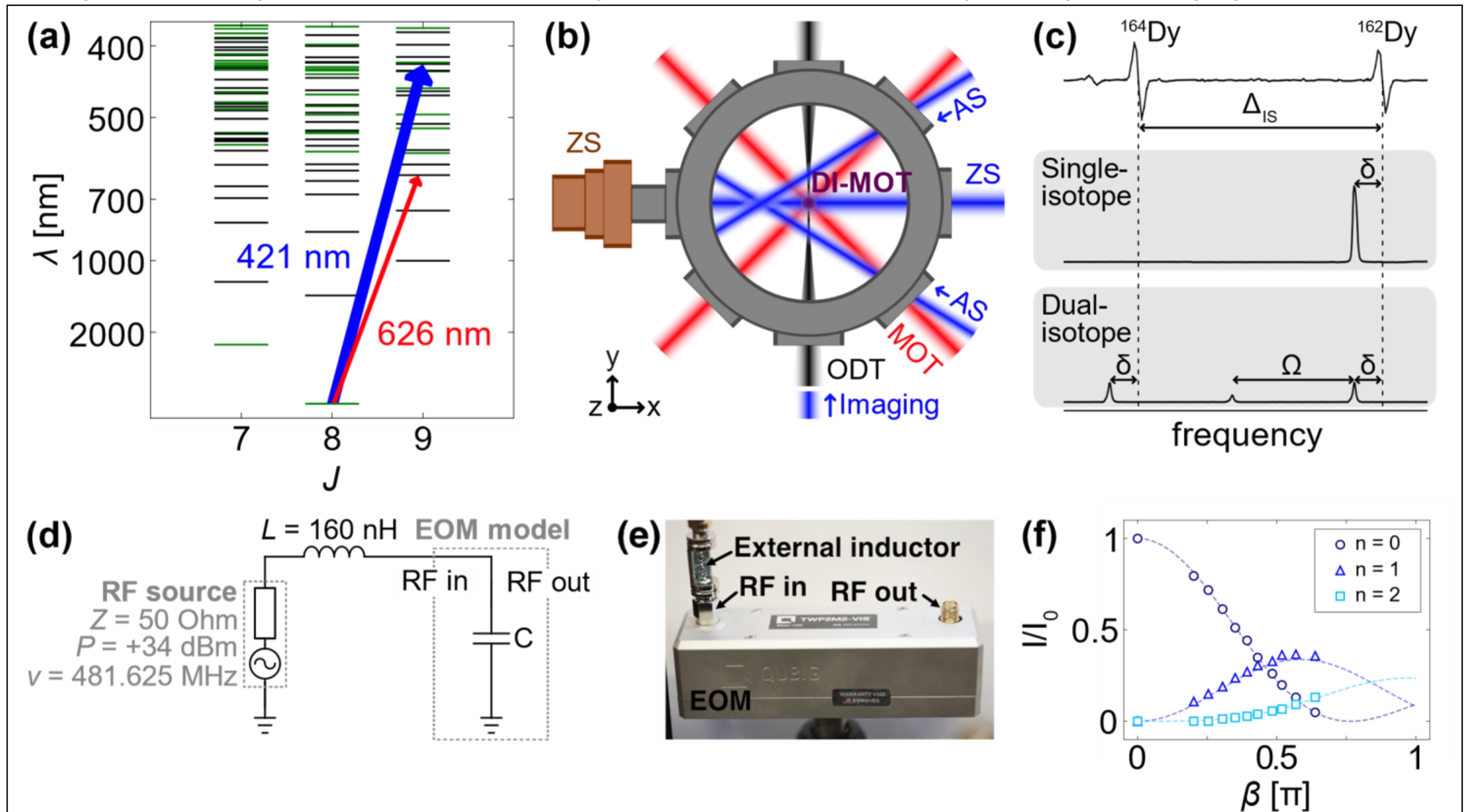

**Figure 1.** a) Dy simplified electronic structure and relevant laser cooling transitions at 421 nm (blue arrow) and 626 nm (red arrow). b) Top view of the science chamber with relevant slowing and trapping laser beams. c) Top: Doppler-free spectroscopy of Dy on the 626-nm transition. Centre: Cavity spectrum of the 626-nm laser in single-isotope operation. The laser frequency is chosen to be red-detuned from the resonance by a frequency $\delta$. Bottom: Cavity-spectrum of the 626-nm laser in dual-isotope operation. Phase modulation in an EOM creates frequency sidebands separated by the drive frequency $\Omega$. Here, we choose $\Omega = \frac{\Delta_{IS}}{2}$ such that the first-order sidebands address each isotope. d) Scheme for the resonant circuit consisting of the non-resonant EOM, modelled with a capacitor C and an inductor L. e) Photo of the EOM with plug-and-play electronic circuit. f) Relative optical power in the different sideband orders $n$ as a function of modulation depth $\beta$, imprinted on a 500-mW laser beam at 626 nm.

## 3. Sideband-MOT

In this section, we characterise the performance of our sideband-MOT, that is, the MOT entirely realised by first-order sidebands in the 626-nm laser. When using sinusoidal phase-modulation, many frequency components, spaced by the EOM-drive frequency, coexist in the laser beams. The Dy narrow-line MOT on the 626-nm transition generally operates around a laser detuning $|\delta| \ll \Delta_{IS}$ and comparable to $\Gamma_{626}$. This condition represents a significant advantage in the realisation of a DI-MOT on a narrow transition, since we do not expect the sidebands at large detuning to introduce scattering losses at the MOT position.
We initially test the performance of the sideband-MOT trapping one isotope at the time. We do not drive the EOM on the 421-nm laser, resulting in the slowing of either $^{162}$Dy or $^{164}$Dy. From the 626-nm laser setup, we deliver to the MOT chamber up to 300 mW of light, which already passed through the non-resonant EOM and an additional low-frequency EOM for spectral broadening[23]. The MOT beams have a diameter of 2 cm and are arranged in a standard, retro-reflected, three-axis configuration, with the MOT coils placed along the gravity direction. The unmodulated 626-nm laser is locked around the centre frequency between the transitions of the two isotopes (see Fig. 1c-bottom). At the maximum optical power, if we do not drive the 626-nm EOM, no atom is trapped. Keeping the optical power constant, we start to trap atoms only as we increase the EOM RF-drive. Driving at half the isotope shift, we ensure to have an equal detuning during all laser cooling stages for both isotopes.

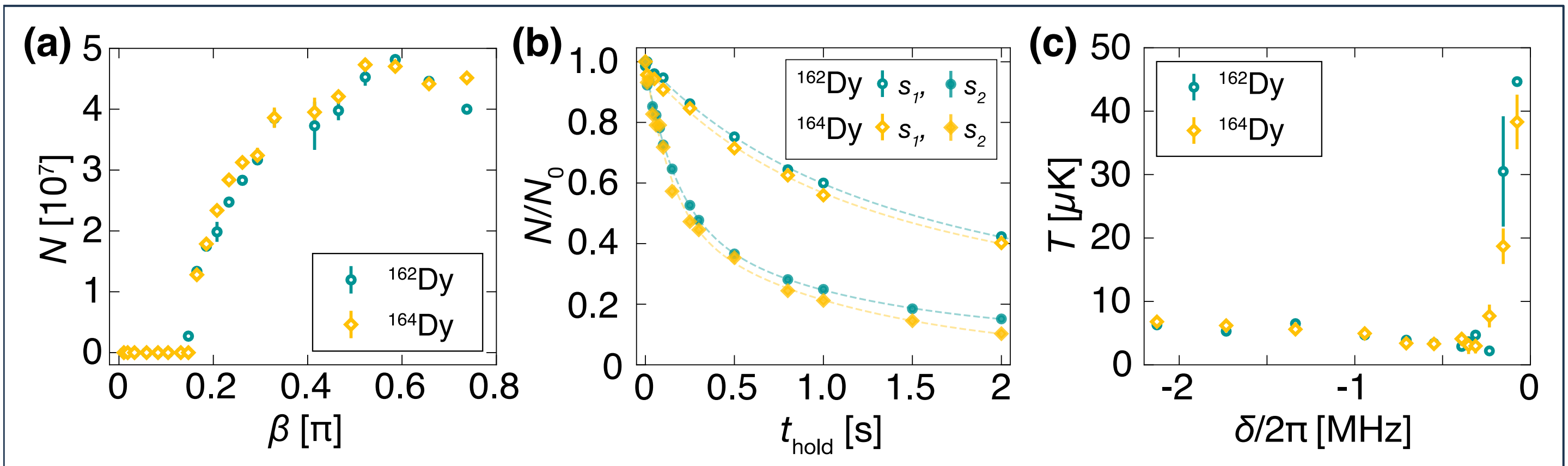

**Figure 2.** Properties of the single-isotope sideband-MOT. a) Number of trapped atoms $N$ in the sideband MOT as a function of $\beta$ at fixed optical power for $^{162}$Dy (green) and $^{164}$Dy (yellow). b) Evolution of $N$ normalised to initial atom number $N_0 = 5 \cdot 10^7$ at different hold times $t_{hold}$ in the compressed sideband-MOT for $^{162}$Dy (green) and $^{164}$Dy (yellow) at a detuning $\frac{\delta}{2\pi} = -1.3$ MHz saturation parameters $s_1 = 0.24$ (hollow) and $s_2 = 0.02$ (solid). c) Temperature of $^{162}$Dy (green) and $^{164}$Dy (yellow) in the sideband-MOT as a function of detuning at $s = 0.02$ .

As shown in Fig. 2a, the number of trapped atoms $N$ in the DI-MOT equally increases with the modulation depth $\beta$ for the two isotopes. $N$ peaks at a modulation depth $\beta_{\max}$, corresponding to the first-order sideband maximum. For higher values of $\beta$, $N$ decreases. The number of trapped atoms in the sideband-MOT is equal to the standard MOT when the optical power in the sidebands matches the unmodulated MOT power. Since the two isotopes experience the same laser detuning and intensity, they show similar dynamics during the compression phase, which we use to increase phase-space density after MOT loading. Conveniently, the same experimental ramps in magnetic field, laser intensity, and detuning $\delta$, cool both isotopes to temperatures of around 10 μK (see Fig. 2c). Additionally, both isotopes show a similar lifetime (see Fig. 2b), which is found to be consistent with previously reported Dy MOTs[27,28]. We conclude that the presence of different sidebands spaced by multiples of $\Delta_{IS}$ does not affect the performance of the sideband-MOT with respect to the standard one.

## 4. Simultaneous slowing of two isotopes

To turn the sideband-MOT into a DI-MOT, we implement a similar strategy on the 421-nm slowing transition. Hence, we introduce sidebands on the 421-nm laser setup to simultaneously slow both isotopes. In this case, we drive a resonant EOM (model QUBIG PM-Dy_0.5) with central frequency $\Omega = \Delta_{\mathrm{IS}}^{421}/2 = 456.5$ MHz. Driving the EOM with a RF-power of 0.5 W is sufficient to transfer most of the optical power to the two first-order sidebands. Similar to the 626-nm case, we move the laser lock point by the drive frequency and obtain the detuning conditions for simultaneously slowing both isotopes. To independently image the isotopes, we pick around 35 mW of light before the resonant EOM and change its frequency with a combination of two AOMs both in double-pass configuration.
In the single-isotope slowing, we operate with 60 mW in the Zeeman Slower (ZS) beam, 300 mW in the TC, and 5 mW in each arm of the AS. To test the dual-isotope slowing, we start without changing the overall laser power and drive the 421-nm EOM at $\beta_{\mathrm{max}}$. We immediately trap in the DI-MOT around 2·10$^6$ atoms for both $^{162}$Dy and $^{164}$Dy. The lower atom number comes from the effective lower power in each sideband. To compensate for this reduction, we increase the overall laser power. In Fig. 3a, the dual-isotope AS shows a maximum efficiency at a power of 15 mW per arm, consistent with the single-isotope configuration. The optimal AS settings result in a factor of 20 improvement in the DI-MOT atom number and loading rate similar to already reported results in the single-isotope case[29]. We increase the power also along the TC path to its maximum value of around 400 mW. In Fig. 3c, we show that the gain from the TC in the number of trapped atoms in the DI-MOT is identical to the single-isotope MOT at the same effective power. We observe that the TC increases the MOT loading rate by a factor of 6 in the dual-isotope case, as opposed to a factor of 8 measured in the single-isotope case. The reduction in the loading rate comes from the limited power available on the TC arms. In both the AS and TC, the first-order sidebands have an optimal detuning of around $\delta \simeq -\Gamma_{421}$ , while the other sidebands are significantly more detuned and do not cause any measurable effect.
The presence of spurious sideband orders has instead a more detrimental impact on the ZS efficiency. The ZS field is designed for a laser detuning $\delta_{ZS} = -2\pi \cdot 700\ \mathrm{MHz} \simeq -22\,\Gamma_{421}$. Since $|\delta_{ZS}| > \Omega$, the EOM modulation introduces sidebands closer to the isotopes' resonances. For instance, at a modulation depth $\beta_{max}$, around 10% of the power is left in the zero-order sideband. Since $^{164}$Dy is the isotope with the lowest transition frequency, the zero-order has a detuning $\delta \simeq -7.5\,\Gamma_{421}$ with respect to the $^{164}$Dy resonance. From a numerical simulation of the ZS dynamics (see Fig. 3b), we expect that the zero-order sideband can over-decelerate the atom class that exits effusion cell with a velocity below 250 m/s. Assuming a Maxwell-Boltzmann distribution of the velocities in the hot Dy beam, we estimate a reduction of 30% of the atom flux at the DI-MOT position. In principle, we can increase the modulation depth to $\beta = 0.75\pi$, lower the zero-order power, and suppress the related losses on $^{164}$Dy. However, in doing so, the second-order sidebands acquire more optical power. Owing to the sidebands' regular spacing, the high-frequency, second-order sideband has an equal detuning $\delta \simeq -7.5\,\Gamma_{421}$ with respect to the $^{162}$Dy resonance. This effect results in similar losses of $^{162}$Dy in the DI-MOT. In Fig. 3d, we show how the control over the relative intensity of the 421-nm sideband affects $N$ in an isotope-dependent way. For these measurements, we increase the overall power in the ZS beam to 130 mW. Without modulation, all the power is in the zero-order and no atom can be decelerated or trapped in the DI-MOT. As $\beta$ is increased, optical power is quickly transferred to the first-orders, but the zero-order retains significantly more power than the second-order. As a result, $^{164}$Dy experiences a stronger over-deceleration and we preferentially trap $^{162}$Dy. The number of trapped $^{162}$Dy peaks around $3.5 \cdot 10^7$ atoms close to $\beta = 0.4\,\pi$, where roughly 45 mW of power are in each first-order sideband. Increasing $\beta$ even further, the effects of the over-deceleration are transferred from one isotope to the other. At around $\beta_{max}$, the number of trapped $^{162}$Dy and $^{164}$Dy is roughly equal, with around $2.5 \cdot 10^7$ atoms for each isotope. At higher $\beta$, $^{164}$Dy reaches a maximum number of $3.7 \cdot 10^7$ atoms, while the number of $^{162}$Dy keeps falling. The control over the EOM modulation depth on the 421-nm laser path grants us a very simple way to control the relative population imbalance in the DI-MOT, which can be convenient for the study of ultracold mixtures and polaron physics[30], once the atoms are transferred into an optical dipole trap.
The lower effective power in TC and the multi-sideband effects in the ZS result in a reduction of almost a factor of 2 in $N$ with respect to the single isotope-case. Conveniently, the reduction can be entirely compensated by increasing the effusion cell temperature by just 50 °C. To the best of our knowledge, our results show the first EOM-based scheme for simultaneous Zeeman slowing of an isotopic mixture into a DI-MOT.

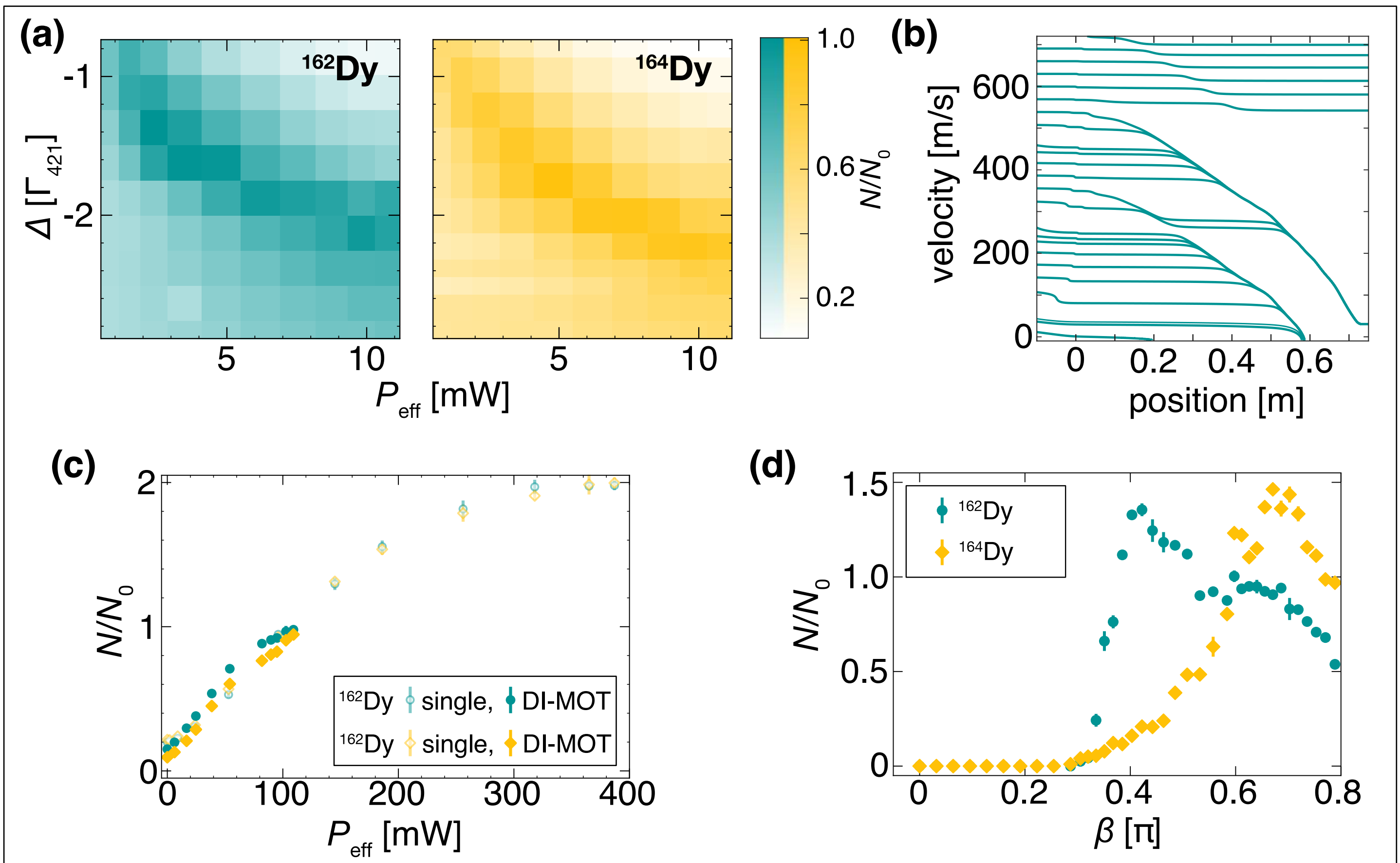

**Figure 3.** Characterisation of the DI-MOT. a) Relative atom number $N/N_0$ in the DI-MOT for both $^{162}$Dy (green) and $^{164}$Dy (yellow) as a function of AS power and detuning. $N_0 = 2.5 \cdot 10^7$ for both isotopes. b) Numerical simulation of the ZS dynamics of $^{164}$Dy for different velocity classes at $\beta = 0.5\pi$. c) Atom number $N$ as a function of effective TC power for single-isotope (hollow) and dual-isotope (solid), d) $N$ as a function of $\beta$ of the 421-nm EOM.

## 5. RF-control of isotopes in the DI-MOT

A characteristic feature of narrow-line MOTs is the interplay among gravity and radiation forces, which results in the MOT equilibrium position being lower than the centre of the MOT quadrupole field. Generally, the MOT position can be finely adjusted by changing the laser detuning, the laser intensity, and the MOT magnetic-field gradient. In our case, detuning the EOM drive from half the isotope shift results in an isotope-dependent detuning of the first-order sidebands. The key advantage is that by changing the EOM frequency by few hundreds of kHz, the vertical MOT position shifts by hundreds of micrometres at the end of the compression phase. Such a small change in EOM drive does not affect the overall efficiency of the DI-MOT. All the MOT parameters such as loading efficiency, cloud size, and final temperature remain unaffected.
Absorption pictures of the isotopes are shown in Fig. 4a. Even for an EOM drive detuning $\delta_{\mathrm{EOM}} = \Omega - \frac{\Delta_{\mathrm{IS}}}{2}$ of the order of few units of $\Gamma_{626}$, the two isotope clouds are spatially displaced along the gravity axis. For $\delta_{\mathrm{EOM}} < 0$ the low-frequency, first-order sideband has a smaller detuning with respect to $^{164}$Dy than the high-frequency one to $^{162}$Dy. This condition results in a stronger radiative force on $^{164}$Dy and a higher equilibrium position along the gravity axis. For $\delta_{\mathrm{EOM}} > 0$, the situation is inverted and $^{164}$Dy sags below $^{162}$Dy.
Figure 4b shows that by tracking the centre-of-mass (CoM) of the two isotopes at different EOM frequencies, we estimate that the two CoM cross at a frequency of 481.63(3) MHz. Neglecting shifts in the equilibrium position due to the small mass difference between the isotopes, we extract $\Delta_{\mathrm{IS}} = 962(1)\,\mathrm{MHz}$, which is consistent with previous spectroscopy measurements[26]. With a more rigorous treatment of the MOT dynamics and equilibrium position, it may be possible to turn our technique into an alternative approach for measuring the isotope shift on narrow laser-cooling transitions.

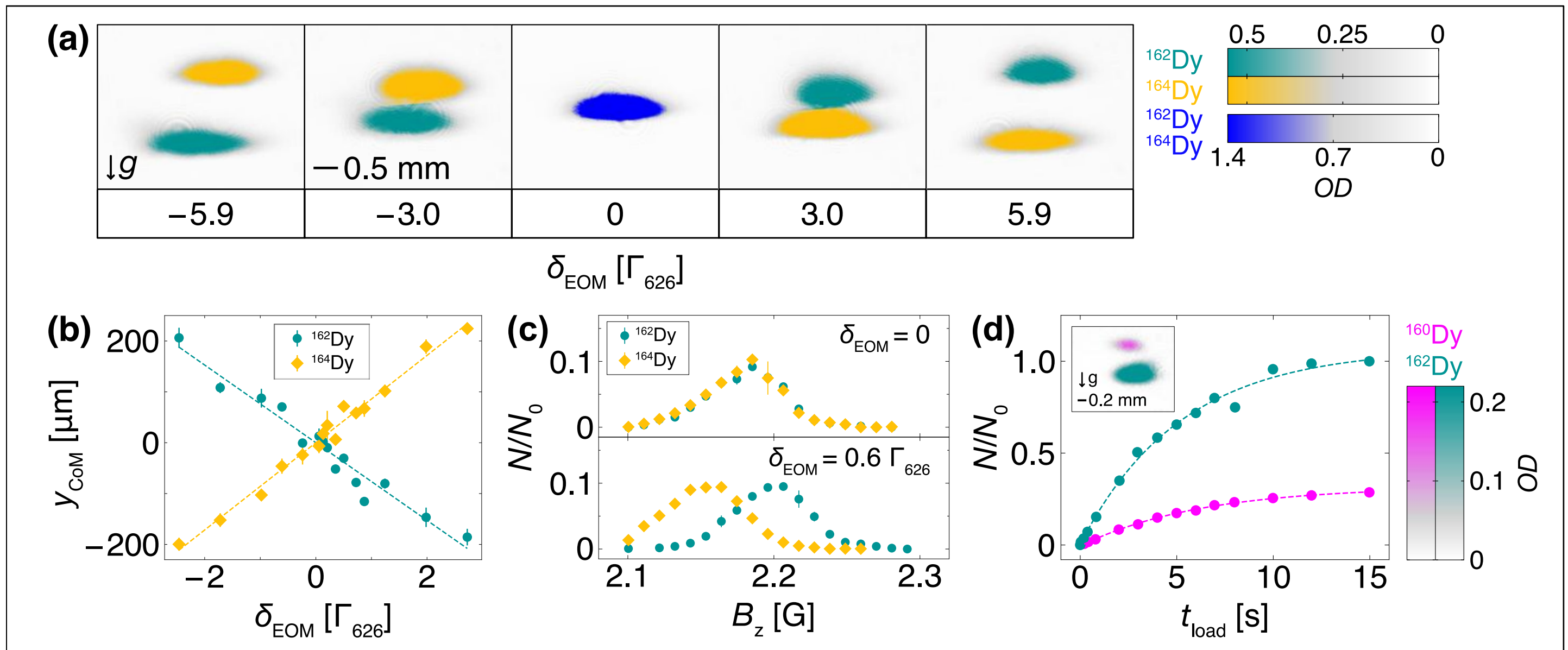

**Figure 4.** a) Optical density (OD) images showing equilibrium positions at different EOM detuning $\delta_{EOM} = \Omega - \frac{\Delta_{IS}}{2}$. b) CoM on the y-axis ($y_c$) as a function of $\delta_{EOM}$. c) Relative atom number $N/N_0$ loaded into the ODT at different $\delta_{EOM}$, as a function of the vertical magnetic field $B_z$. d) DI-MOT loading for $^{160}$Dy - $^{162}$Dy mixture. Inlet shows an OD image of the $^{160}$Dy - $^{162}$Dy DI-MOT at $\delta_{EOM} = 5\Gamma_{626}$.

Finally, we test the loading efficiency of the isotopes into an optical dipole trap (ODT). In our experiment, we use a single-beam ODT derived from a 50-W fibre-amplifier laser at 1064 nm (see Fig. 1b). At the DI-MOT position, the ODT beam has a cylindrical waist of around 25 $\mu$m and polarization perpendicular to the gravity axis. To adjust the overlap among the clouds and the ODT, we scan the offset magnetic field $B_z$. Conveniently, even for $\delta_{\mathrm{EOM}} \neq 0$, the isotopes have similar temperatures and densities. Hence, as depicted in Fig. 4c, we observe a similar maximum loading efficiency of 10% into the ODT. Playing with $\delta_{\mathrm{EOM}}$ provides a simple control knob over the isotope ratio loaded into the ODT, which is convenient for controlling the population imbalance at ultracold temperatures.
To show the modularity of our approach in addressing different isotopic mixtures, we replace the inductor *L* at the 626-nm EOM to have a resonance frequency at 513.5 MHz, that is, half the isotope shift between $^{160}$Dy and $^{162}$Dy on the 626-nm transition[26]. No particular adjustment is needed on the 421-nm EOM since the $^{160}$Dy - $^{162}$Dy isotope shift of 481.8 MHz on this transition is within the bandwidth of the EOM. By simply adjusting the laser lock-point, our scheme immediately produces a DI-MOT of $^{160}$Dy - $^{162}$Dy, as shown in Fig. 4d. The lower natural abundance of $^{160}$Dy results in a lower atom number in the DI-MOT for this isotope, while the temperature and cloud size show an identical dynamic to the other isotope.

## 6. Conclusion

We have shown a simple scheme that exploits the first-order sidebands generated by EOMs to simultaneously slow and trap different isotopes of Dy into a narrow-line DI-MOT. To transfer sufficient optical power into the first-order sidebands, we have used in-house designed, plug-and-play electronics elements that turn a broadband EOM into a resonant circuit. By tuning the sideband spacing for both the 421- and 626-nm transitions, we realised DI-MOTs of $^{162}$Dy - $^{164}$Dy and $^{160}$Dy - $^{162}$Dy.
The $^{162}$Dy - $^{164}$Dy DI-MOT exhibits nearly identical cooling characteristics throughout all laser cooling stages, which is encouraging for subsequent studies into an ODT.
A detailed characterization of these stages revealed, that the trapped atom number represents the biggest difference with respect to the single-isotope case showing a reduction of approximately a factor of 2. The reduction could be attributed to the combined effects of the presence of spurious sidebands in the ZS beam and a lower effective power of the TC beams. Notably, this reduction is less severe than those reported in previous DI-MOTs employing sequential loading schemes, while maintaining a balanced loading ratio. Furthermore, we benefit from a reduced overall sequence time and an equal DI-MOT loading efficiency for both isotopes. In the future, by replacing the ZS with a 2D-MOT[31], it may be possible to completely remove the detrimental effects of the sidebands and improve the trapped atom number.
More generally, our approach can be extended to laser cooling fermionic isotopes as well, or even other atomic species, including Er, ytterbium, and strontium that feature repumper-free cooling schemes and similar isotope shifts to Dy.
Beyond enabling simultaneous cooling, the RF-controlled displacement of the isotopes in the DI-MOT may turn beneficial for sequent stages of the experiment involving optical trapping. In the context of single-atom trapping, the vertical displacement can allow a simpler formation of dual-element arrays, which offer benefits for quantum computing schemes. Additionally, we can easily control the population imbalance of ultracold mixtures in an ODT, which is convenient for studying molecule formation[32] and impurity physics in quantum degenerate gases[30]. These features make the proposed DI-MOT scheme a practical platform for future studies of ultracold isotope mixtures, single-atom trapping and quantum many-body physics.

ACKNOWLEDGMENTS
We thank H. Haak for the mechanical design of our apparatus, and S. Kray, R. Thomas, D. Fontoura Barroso and the members of the electronic and mechanical workshops at FHI for their valuable technical assistance. G.V. acknowledges support from the European Union (ERC, LIRICO 101115996).

## References

1. Schreck, F. & Druten, K. van. Laser cooling for quantum gases. *Nat. Phys.* **17**, 1296–1304 (2021).

2. Kogel, F., Garg, T., Rockenhäuser, M., Morales-Ramírez, S. A. & Langen, T. Molecular laser cooling using serrodynes: implementation, characterization and prospects. *New J. Phys.* **27**, 055001 (2025).
3. Holland, C. M., Lu, Y. & Cheuk, L. W. Synthesizing optical spectra using computer-generated holography techniques. *New J. Phys.* **23**, 033028 (2021).
4. Chomaz, L. *et al.* Dipolar physics: a review of experiments with magnetic quantum gases. *Rep. Prog. Phys.* **86**, 026401 (2022).
5. Lu, M., Burdick, N. Q., Youn, S. H. & Lev, B. L. Strongly Dipolar Bose-Einstein Condensate of Dysprosium. *Phys. Rev. Lett.* **107**, 190401 (2011).
6. Tanzi, L. *et al.* Observation of a Dipolar Quantum Gas with Metastable Supersolid Properties. *Phys. Rev. Lett.* **122**, 130405 (2019).
7. Böttcher, F. *et al.* Transient Supersolid Properties in an Array of Dipolar Quantum Droplets. *Phys. Rev. X* **9**, 011051 (2019).
8. Chomaz, L. *et al.* Long-Lived and Transient Supersolid Behaviors in Dipolar Quantum Gases. *Phys. Rev. X* **9**, 021012 (2019).
9. Böttcher, F. *et al.* New states of matter with fine-tuned interactions: quantum droplets and dipolar supersolids. *Rep. Prog. Phys.* **84**, 012403 (2021).
10. Trautmann, A. *et al.* Dipolar Quantum Mixtures of Erbium and Dysprosium Atoms. *Phys. Rev. Lett.* **121**, 213601 (2018).
11. Bisset, R. N., Ardila, L. A. P. & Santos, L. Quantum Droplets of Dipolar Mixtures. *Phys. Rev. Lett.* **126**, 025301 (2021).
12. Scheiermann, D., Ardila, L. A. P., Bland, T., Bisset, R. N. & Santos, L. Catalyzation of supersolidity in binary dipolar condensates. *Phys. Rev. A* **107**, L021302 (2023).
13. Kirkby, W. *et al.* Excitations of a Binary Dipolar Supersolid. *Phys. Rev. Lett.* **133**, 103401 (2024).
14. Süptitz, W., Wokurka, G., Strauch, F., Kohns, P. & Ertmer, W. Simultaneous cooling and trapping of $^{85}$Rb and $^{87}$Rb in a magneto-optical trap. *Opt. Lett.* **19**, 1571–1573 (1994).
15. Prevedelli, M. *et al.* Trapping and cooling of potassium isotopes in a double-magneto-optical-trap apparatus. *Phys. Rev. A* **59**, 886–888 (1999).

16. Xu, X., Loftus, T. H., Hall, J. L., Gallagher, A. & Ye, J. Cooling and trapping of atomic strontium. *JOSA B 20, 968-976* (2003).
17. Poli, N. *et al.* Cooling and trapping of ultracold strontium isotopic mixtures. *Phys. Rev. A* **71**, 061403 (2005).
18. Chicireanu, R. *et al.* Simultaneous magneto-optical trapping of bosonic and fermionic chromium atoms. *Phys. Rev. A* **73**, 053406 (2006).
19. Stas, R. J. W., McNamara, J. M., Hogervorst, W. & Vassen, W. Simultaneous Magneto-Optical Trapping of a Boson-Fermion Mixture of Metastable Helium Atoms. *Phys. Rev. Lett.* **93**, 053001 (2004).
20. Lu, M., Burdick, N. Q. & Lev, B. L. Quantum Degenerate Dipolar Fermi Gas. *Phys Rev Lett* **108**, 215301 (2012).
21. Valenzuela, V. M., Hamzeloui, S., Gutiérrez, M. & Gomez, E. Multiple isotope magneto-optical trap from a single diode laser. *JOSA B* **30**, 1205–1210 (2013).
22. Knöckel, H. & Tiemann, E. IodineSpec5. Institut für Quantenoptik, Universität Hannover: Hanover, Germany (2013).
23. Maier, T., Kadau, H., Schmitt, M., Griesmaier, A. & Pfau, T. Narrow-line magneto-optical trap for dysprosium atoms. *Opt. Lett.* **39**, 3138–3141 (2014).
24. Lucioni, E. *et al.* Dysprosium dipolar Bose-Einstein condensate with broad Feshbach resonances. *Phys. Rev. A* **97**, 060701 (2018).
25. Leefer, N., Cingöz, A. & Budker, D. Measurement of hyperfine structure and isotope shifts in the Dy 421 nm transition. *Opt. Lett.* **34**, 2548–2550 (2009).
26. Hogervorst, W., Zaal, G. J., Bouma, J. & Blok, J. Isotope shifts and hyperfine structure of natural dysprosium. *Phys. Lett. A* **65**, 220–222 (1978).
27. Dreon, D. *et al.* Optical cooling and trapping of highly magnetic atoms: the benefits of a spontaneous spin polarization. *J. Phys. B At. Mol. Opt. Phys.* **50**, 065005 (2017).
28. Anich, G., Höllrigl, N., Kreyer, M., Grimm, R. & Kirilov, E. Comprehensive characterization of an apparatus for cold electromagnetic dysprosium dipoles. *Phys. Rev. A* **110**, 023311 (2024).

29. Lunden, W. *et al.* Enhancing the capture velocity of a Dy magneto-optical trap with two-stage slowing. *Phys. Rev. A* **101**, 063403 (2020).

30. Scazza, F., Zaccanti, M., Massignan, P., Parish, M. M. & Levinsen, J. Repulsive Fermi and Bose Polarons in Quantum Gases. *Atoms* **10**, 55 (2022).

31. Jin, S. Two-dimensional magneto-optical trap of dysprosium atoms as a compact source for efficient loading of a narrow-line three-dimensional magneto-optical trap. *Phys. Rev. A* **108**, (2023).

32. De Marco, L. *et al.* A degenerate Fermi gas of polar molecules. *Science* **363**, 853–856 (2019).